# Defect-tolerant electron and defect-sensitive phonon transport in quasi-2D conjugated coordination polymers

Hio-Ieng Un[1]*, Kamil Iwanowski[2], Jordi Ferrer Orri[1,5], Ian E. Jacobs[1,5], Naoya Fukui[3], David Cornil[4], David Beljonne[4], Michele Simoncelli[2]*, Hiroshi Nishihara[3], Henning Sirringhaus[1]*

[1] Optoelectronics Group, Cavendish Laboratory, University of Cambridge, JJ Thomson Avenue, Cambridge CB3 0HE, UK

[2] Theory of Condensed Matter Group, Cavendish Laboratory, University of Cambridge, JJ Thomson Avenue, Cambridge CB3 0HE, UK

[3] Research Institute for Science and Technology, Tokyo University of Science, Noda-shi, Chiba 278-8510, Japan

[4] Laboratory for Chemistry of Novel Materials, University of Mons, Mons, Belgium

[5] These authors contribute equally to this work

* Corresponding authors, email: hiu20@cam.ac.uk, ms2855@cam.ac.uk, hs220@cam.ac.uk

## Abstract

Thermoelectric materials, enabling direct waste-heat to electricity conversion, need to be highly electrically conducting while simultaneously thermally insulating. This is fundamentally challenging since electrical and thermal conduction are usually coupled. Here, we discover that quasi-2D conjugated coordination polymer films exhibit this ideal mix of antithetical properties due to coexistence of defect-tolerant charge transport and defect-sensitive heat propagation. The former is highlighted by the highest conductivities > 2000 S cm$^{-1}$ with metallic temperature dependence observed in disordered films with paracrystallinity > 10%, while the latter manifests in low, temperature-activated lattice thermal conductivities (< 0.38 W m$^{-1}$ K$^{-1}$) originating from small-amplitude, quasi-harmonic lattice dynamics with disorder-limited lifetimes and vibrational scattering length on the order of interatomic spacing. Based on temperature-dependent thermoelectric and magnetotransport experiments we identify a two-carrier (hole-electron), ambipolar metallic transport regime as the origin of relatively small Seebeck coefficients in these materials. Our findings identify conjugated coordination polymers as attractive materials for applications in thermoelectric energy harvesting, (bio)electronics and energy storage.

# Main

Structure determines properties and uses of materials. Electron and phonon motion as well as energy transfer processes are often faster and more efficient in crystalline materials than in amorphous solids. Single or poly-crystalline silicon have carrier mobilities three orders of magnitude higher than amorphous Si[1], and graphene has electrical (thermal) conductivity ten (three) orders of magnitude higher than glassy carbon[2,3]. To block the heat flow without significantly degrading the electronic transport, traditional high-performing inorganic thermoelectric materials have been developed by introducing a large mass fluctuation into crystalline structure as point scatterers and meanwhile reducing grain size below phonon but above carrier mean free paths[4]. However, difficult trade-offs remain; often, the lattice thermal conductivity remains well above the amorphous limit[5]. Not only to maintain incremental performance but also to address scientific and technological challenges that require disruptive breakthroughs, finding new materials with fast charge but slow heat transfer in a single structure is therefore fundamentally challenging but required.

Conjugated metal-organic frameworks (cMOFs) or conjugated coordination polymers (cCPs)[6–8], which represent a new frontier in (semi)conducting materials combining some unique electronic properties of inorganic materials such as high mobility and the chemical tunability and ease of processing of organic semiconductors, have been theoretically speculated to have interesting thermoelectric properties[9–11]. These speculations have been validated in our recent work[12] where we experimentally demonstrated that a family of Ni-based cCPs can reach promising level of thermoelectric performance with near metallic charge transport, notably, in poorly crystalline films. Compared to Ni-based cCPs, copper benzenehexathiol ($Cu_3BHT$) exhibits higher electrical conductivity and is the only cCP discovered to be superconducting with a transition temperature of 0.2 K[13]. The electrical conductivity values of $Cu_3BHT$ reported to date, span over a large range, from close to zero to 2500 S cm$^{-1}$, across several studies using the same synthesis method (liquid-liquid interfacial synthesis)[13–16]. The chemistry behind the films synthesized with such large variations in electrical conductivity remains unknown.

The main objective of this work is to study the structure – thermoelectric properties relationship of thin films of $Cu_3BHT$ as a model system with controlled degrees of chemical and structural defects. We resolve and quantify the chemical and structural defects at (sub)micrometer length scales, and directly visualize the structural orientation and distribution of grains by scanning electron diffraction (SED) at the nanoscale. By relating our structural measurements of the thermoelectric and magneto-transport coefficients, we discover that $Cu_3BHT$ is an unconventional material, in which electron motion in chemically defective, poorly crystalline films with paracrystallinity > 10% can be delocalized and metallic. However, lattice vibrations are rather localized and incoherent, leading to a reduction of the lattice thermal conductivity to minimum theoretical values. In this way it becomes possible to disrupt heat transport not just without affecting, but even enhancing electronic transport. This phenomenon is observed neither in traditional inorganic or organic semiconductors, and implies a new advantageous thermoelectric transport regime in non-crystalline coordination materials that is distinct from the classical intrinsic phonon-glass, electron-crystal (PGEC) concept that can be realized in some highly crystalline inorganic materials.

## Quantification of Chemical and Structural Disorder

In the liquid-liquid interfacial synthesis of our Cu-BHT films we vary the molar ratio between the Cu precursor and the BHT in the growth solution from 2 to 7 around the nominally ideal ratio of 3 for perfect $Cu_3BHT$ (Fig. 1a, see Methods for details). Hereafter the ratio we added in the synthesis is referred to as the Cu/BHT ratio. In this way we aim to control the level of BHT and copper vacancies in the films and study the effects of the associated defects on the thermoelectric coefficients. $Cu_3BHT$ has been considered to adopt a two-dimensional (2D) layered structure[13–16]. However, there is a very recent work[17] reporting a new, non-van der Waals layered structure for $Cu_3BHT$, which is resolved by more precise structural determination with atomic resolution. As a result we employ the new, non-van der Waals layered structure in this work, which is also justified by our simulation (more on this later). Grazing-incidence wide-angle X-ray scattering (GIWAXS) images of the most crystalline films obtained from a Cu/BHT ratio of 2 (Fig. 1b) show three well-defined diffraction peaks attributed to the intra-sheet periodicity ($h$00) in the in-plane direction and an arc diffraction corresponding to the π-stacking diffraction (00$l$) in the out-of-plane direction, indicating an anisotropic and face-on-preferred orientation. Films grown with higher Cu/BHT ratio are less crystalline and have a more amorphous microstructure (Fig. 1c, Supplementary Fig. 6b-d), that is also apparent from the corresponding secondary electron microscopy (SEM) images. Scanning electron diffraction (SED) in transmission mode was employed to directly visualize the stacking orientation on the nanoscale (Fig. 1f,g, Supplementary Fig. 7). By acquiring and indexing each diffraction pattern in the scanned area an orientation map can be computed. A face-on stacking preference with the [002] zone axis normal to substrate (indicated in red), is seen to dominate in both Cu/BHT ratios of 2 and 3.5, while edge-on preference, i.e. zone axes [200] and [020] normal to substrate (green and blue, respectively), are less prominent. The black areas in Fig. 1f,g show scanned regions where the diffraction signal was not good enough to estimate the zone axes, presumably indicative of disordered or amorphous regions as well as grain boundaries; this result suggests that in addition to ordered grain boundaries, crystalline regions are possibly separated by more disordered/amorphous regions. Energy dispersive X-ray spectroscopy (EDX) was performed to quantify the chemical composition (Supplementary Fig. 3): The films were found to consist of Cu, S, C, and unexpectedly, some O. Even when an excess of organic ligand was added, the atomic Cu-to-S ratio is higher than the ideal value of 0.5. Under all growth conditions used the composition of the films remains Cu-rich, and the Cu/S ratio is positively correlated with the Cu/BHT ratio used during film growth (Fig. 1d). These results suggest that the majority of chemical defects are BHT vacancies, which may be filled with the acetate anions of the metal precursor ($CH_3COO^−$). Accordingly, a gradually increased coordination imperfection is observed in the Raman spectra (Fig. 1e): with increasing Cu/BHT ratio a broad band centered at ca. 90 cm$^{−1}$, which is mainly attributed to Cu displacement, vanishes and a well-resolved triplet band at 305.0, 328.1, and 357.9 cm$^{−1}$, which is due to S displacement, loses its fingerprint and merges into a broad band as we move from crystalline to more amorphous films. This is accompanied by a blue shift of 2 – 6.3 cm$^{−1}$ of the Raman vibrational frequency, which suggests the presence of compression-dominated lattice strain variations. Simultaneously, the intra-sheet lattice spacing (200) reduces from relatively ideal 7.38 ± 0.07 (7.3531 Å in crystal structure)[17] to 7.16 ± 0.03 Å as the Cu/BHT ratio increases (Fig. 1b). SED also resolved the lattice parameter variations, indicating grains exhibiting strain can be nanoscale-ordered (Supplementary Fig. 7). Such a lattice parameter variation, as well as strain, are generated by the presence of vacancies and dislocations[18,19]. We note that strain is the more likely explanation as opposed to presence of another phase since X-ray and electron diffraction patterns as well as Raman spectra did not exhibit different patterns or features, but merely small local changes in spacing (~8%) or frequency (< 2 cm$^{−1}$) (Supplementary Sections 2.2 and 2.3).

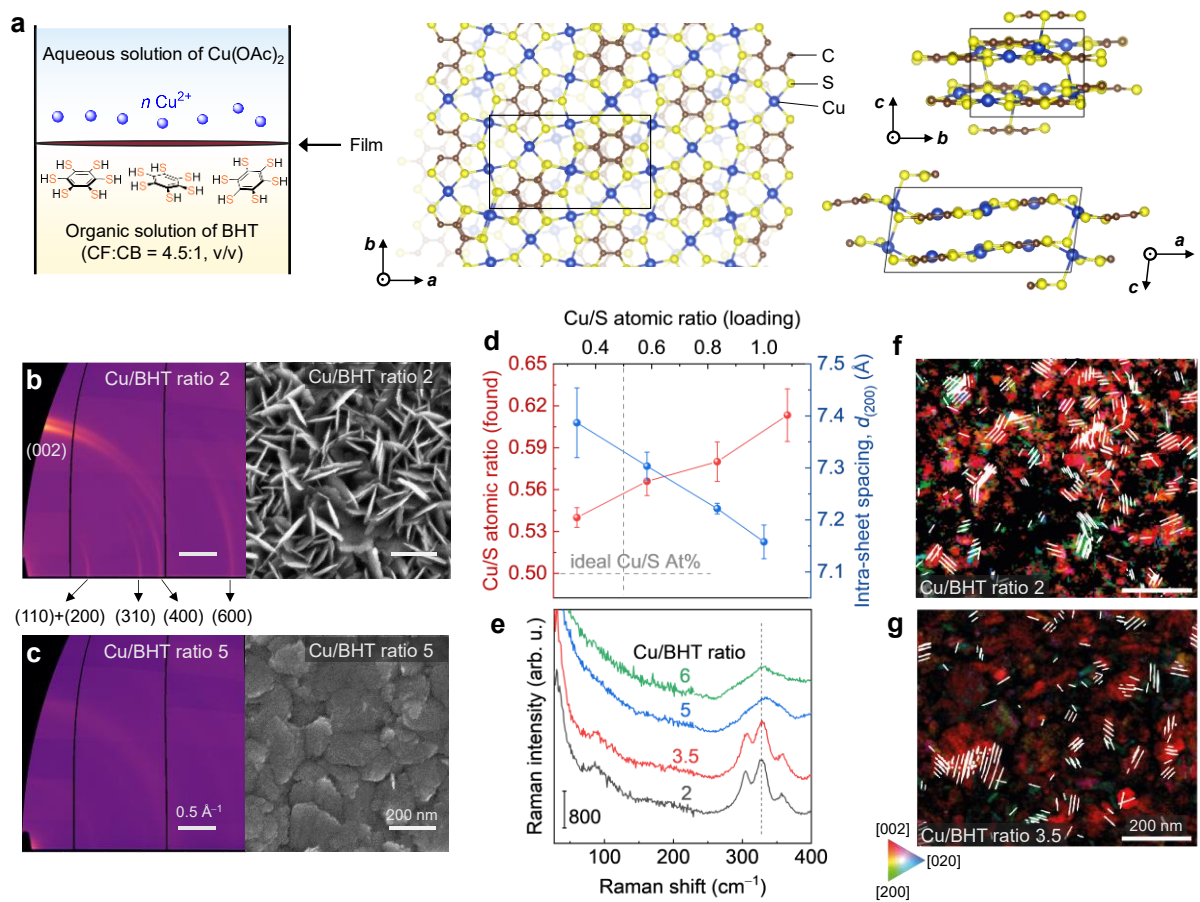

**Figure 1. Quantification of structural disorder and chemical defects in Cu-BHT films as a function of composition. a**, Synthesis of Cu-BHT films by liquid-liquid interfacial method and diagram of the non-defective lattice of a new $Cu_3BHT$ structure reported recently[17]. *n* in the chemical reaction formula is the ratio of $Cu^{2+}$ to BHT added, varying between 2 – 7 in this work. **b, c,** GIWAXS and SEM images for a crystalline sample with Cu/BHT ratio of 2 and a more amorphous sample with a Cu/BHT ratio of 5, respectively. **d**, Cu/S atomic ratios in the films detected by EDX and intra-sheet lattice spacing $d_{(200)}$ characterized by X-ray diffraction as a function of Cu/BHT mole ratio (loading). The error bar for Cu/S atomic ratio represents the standard deviation of the experimental results taken at different local positions in a same sample and that for the spacing $d_{(200)}$ represents the potential error for the fitting of the diffraction peaks from which the spacing $d_{(200)}$ extracted. **e**, Raman spectra of different compositions, showing changes of distinct Raman bands around 90 and 320 $cm^{-1}$ corresponding to mainly Cu atom displacements and mainly S atoms displacements, respectively (see also Fig. 4**b**). **f, g**, SED maps visualizing the stacking orientation of films with Cu/BHT ratio of 2 and ratio of 3.5 on the nanoscale. Face-on dominant grains (zone axis [002] mostly normal to the substrate) are highlighted in red, edge-on dominant grains (zone axes [200] and [020] mostly normal to the substrate) are highlighted in green and in blue, respectively. White streak lines correspond to grains with intermediate/mixed orientation. Black areas show the regions where diffraction is not good enough to estimate the zone axes. Corresponding electron diffraction patterns for face-on and edge-on and the virtual dark-field image can be found in Supplementary Fig. 7.

To better understand the correlation between chemical defects and structural imperfections as well as to quantify the relative contributions from cumulative, paracrystalline disorder, *g,* and from (strain-related) lattice parameter fluctuations, $e_{rms}$, pseudo-Voigt peak shape analysis[20] was applied to a series of in-plane (*h*00) diffraction peaks. The pseudo-Voigt mixing parameter $\eta$, in all samples is found to be closer to 1 (Lorentzian peak shape) than to 0 (Gaussian peak shape) (Fig. 2a), suggesting "long-range" ordering in all compositions is paracrystallinity-dominated, with less important contributions from the strain-related lattice parameter fluctuations[20]. The presence of three orders of the (*h*00) diffractions enables us to perform fast-Fourier-transform (FFT) Warren-Averbach (WA) analysis[20,21] (Fig. 2b, Supplementary Fig. 6i-l), which allows quantifying the contributions of different types of disorder. The X-ray coherence length, $L_c$, i.e. the crystallite size, *g*, and $e_{rms}$ are calculated to be 18.5 ± 1.0 nm, 4.8 ± 1.2 %, and < 1% for the intra-sheet lattice of the most crystalline, ordered Cu/BHT ratio of 2. As the Cu/BHT ratio increases, the diffraction peaks become rings and significantly broaden, with order-dependent broadening being much stronger (Supplementary Fig. 6a-g), falling within the range of strongly disordered materials where coherence is sufficiently hindered by disorder, not finite size. In this regime, a disorder-associated coherence length $\xi$ is used instead of $L_c$[20]. *g* and $\xi$ are found to gradually increase to 13% and reduce to < 10 nm, respectively, as the Cu/BHT ratio increases (Fig. 2c).

Likewise, the parameter for the out-of-plane (00*l*) ordering have similar compositional-dependence. These results reveal that lower Cu/BHT ratio provides higher mean crystallite size and better "long-range" ordering. The transition from a crystalline to a more disordered, quasi-amorphous structure is believed to be induced by chemical defects that distort the lattice.

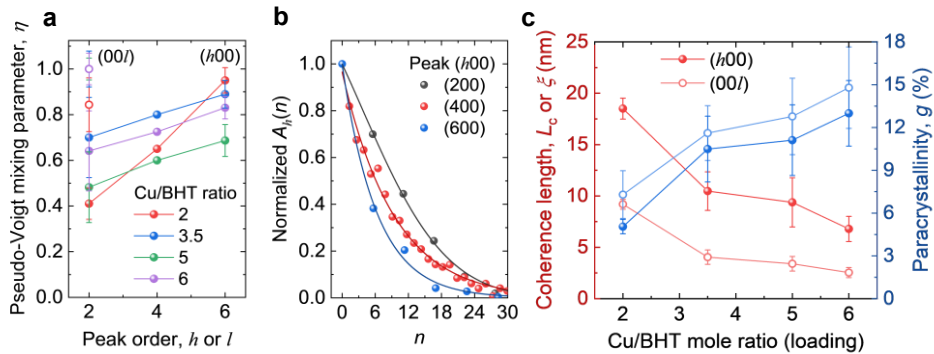

**Figure 2. Quantified analysis of different contributions to structural ordering in Cu-BHT films as a function of composition. a**, Pseudo-Voigt mixing parameter $\eta$, in both (*h*00) (solid circles) and (00*l*) (open circles) obtained by fitting the diffraction profiles with Pseudo-Voigt function. **b**, WA full-fit of isolated peaks (*h*00) of a sample with Cu/BHT ratio of 2 using equation $\ln A_h(n) = \ln \frac{N(n)}{N_3} - 2\pi^2 h^2 n f(n)$, where $f(n) = g^2 + \langle ne^2 \rangle$. By fitting/plotting the Fourier transform coefficients $A_n$ of the diffraction peaks against diffraction order *h* with holding *n* as constant, the term of $-2\pi^2 n f(n)$ can be determined from the slope. Thereafter the effects of paracrystalline disorder, $g$, and lattice parameter fluctuation, $\langle e^2 \rangle$ (also termed as $e_{rms}$), can be further separated by the relation of $f(n)$ and $n$. More details about the analysis method can be found in Supplementary Section 2.2. **c**, X-ray coherence length and paracrystalline disorder evaluated from WA analysis for the in-plane (*h*00) and out-of-plane (00*l*) diffractions as a function of Cu/BHT mole ratio. The error bars in **a**, **c** reflect the uncertainties of the values due to the fit of the diffraction peaks.

**Defect-Manipulated Thermoelectric Transport**

The presence of these chemical and structural defects strongly influences the thermoelectric properties of the films. As the Cu/BHT ratio increases, the thermal conductivity $\kappa$ and electrical conductivity $\sigma$ in a direction parallel to the film plane increase at first and then decrease while the Seebeck coefficient $S$ decreases at first and then increases (Fig. 3a–c). Unexpectedly, the most chemically perfect, crystalline composition (Cu/BHT ratio of 2) does not show the highest electrical conductivity, but only exhibits a value of 636 ± 245 S cm$^{-1}$, though this is still considerably larger than many other 2D conducting coordination nanosheets reported to date[22–24]. Interestingly, more amorphous compositions (i.e. Cu/BHT ratio of 3.5 – 5.5) with paracrystallinity > 10% and less prominent grain boundary features (Figs. 1c and 2d) exhibit electrical conductivities of up to ca. 2000 S cm$^{-1}$. The plot of thermal conductivities versus electrical conductivities of different compositions shows a good linear relationship (Supplementary Fig. 16), from which the Lorenz number $L$ is extracted to be $(2.77 \pm 0.27) \times 10^{-8}$ $V^2$ $K^{-2}$ according to the Wiedemann-Franz (W-F) law. Within the experimental error this matches with the theoretical value of $2.44 \times 10^{-8}$ $V^2$ $K^{-2}$ for Fermi liquids and free-electron metals. The high electrical conductivities, small Seebeck coefficients and the free-electron Lorenz number suggest that the electronic conduction in Cu-BHT is likely in a degenerate regime and possibly has an intrinsically metallic nature.

We also investigated the temperature dependence of these transport coefficients. Increasing chemical and structural defects cause a clear transition (regime of Cu/BHT ratio of 2 to 5 – 5.5) from a weakly thermally activated electrical conductivity to a metallic behavior with conductivity increasing with decreasing temperature (Fig. 3g,h). Correspondingly, the temperature-dependent Seebeck coefficient changes from a superlinear to a near linear temperature dependence behavior (Fig. 3f). A sign change in Seebeck coefficient over temperature, most prominently in Cu/BHT ratio of 4.5, was observed, suggesting the presence of both electron and hole pockets on the Fermi surface. Usually, metallic transport tends to occur in highly ordered crystalline or semicrystalline phase in many inorganic and organic (semi)conductors, while glassy systems with pronounced disorder (paracrystallinity > 10%) often exhibit a strong temperature dependence of the electrical conductivity that drops by orders of magnitude from room to low temperature[1,2]. This is fundamentally inconsistent with our defect-driven metallic transition as well as the drops by only < 20% in electrical conductivities from 300 K to ~25K even in the most disordered films (Fig. 3g,h). It is important to understand the origin of this remarkably disorder-tolerant charge conduction regime. Our attempts to apply transport models for electrons hopping between localized states[25,26] and trapping below extended states[27,28], which have been used to fit the temperature dependent electrical conductivity in other highly conducting materials such as coordination polymer NiTTFtt[29] and organic polymers, such as PBTTT[30] and PBFDO[31] in the (near) degenerate regime, were unsuccessful (Supplementary Section 5.2 and Figs. 17-18).

The observation of distinct grain boundaries by SEM and SED (Fig. 1b,f) suggests that electronic transport across grain boundaries is likely to be the limiting mechanism in the most crystalline films[5]. This mechanism is demonstrated by using a heterogeneous transport model[32] that decomposes the measured electrical conductivity into contributions from intragrain metallic transport pathways and from disordered metallic and hopping transport pathways in the grain boundaries (Fig. 3h,i, Supplementary Section 5.2, Fig. 19). The modelling shows clearly that the electrical conductivity of the crystalline films is indeed grain boundary limited, with the hopping energy between grains found to be 637 K (55 meV), and that the transport through the grain boundaries is supported by the disordered metallic pathway. It is possible that the local potential barriers associated with the grain boundaries could also induce an energy filtering effect (Fig. 3d), which selectively removes the contributions of low-energy charge carriers[5,33] and could potentially be the reason for the increase of the Seebeck coefficient in the most crystalline samples (Fig. 3b,f). In the samples with higher paracrystallinity but less distinct grain boundaries, the energetic landscape becomes smooth and a by ~0.10 ± 0.05 eV deeper Fermi level is generated, as supported by ultraviolet photoelectron spectroscopy (UPS) measurements (Supplementary Fig. 9). The down shifted Fermi level moves the transport away from a hole-electron ambipolar transport regime into a hole dominated regime (as will be further discussed later).

Next, we analysed the lattice thermal conductivities of the different compositions (Fig. 3e), which can be calculated from the measured thermal conductivity by using the W-F law (Supplementary Fig. 16). One of the most intriguing aspects of this material system is that significant structural disorder (paracrystallinity > 10%) enables efficient, metallic charge motion but heat transport carried by lattice vibrations is pushed to the theoretical minimum limit (will be further discussed in the next section) with a mild increase with temperature. Even in the most crystalline composition, the lattice thermal conductivity is only 0.38 W m$^{-1}$ K$^{-1}$. This value is record low for homogeneous, non-superlattice crystalline solids with extended networks of atoms bonded through non-weak interactions; it is two orders of magnitude lower than the typical values of covalent solids[34] (> 10 W m$^{-1}$ K$^{-1}$), and comparable to the conductivity of soft, van der Waals-bonded organic semiconductors[35–37] (< 0.5 W m$^{-1}$ K$^{-1}$). Overall, our measurements in different Cu/BHT compositions highlight the coexistence of metallic, nearly free-electron-like electrical conductivity and record-low, thermally activated heat conductivity, *i.e.,* our disordered Cu-BHT films exhibit *defect-tolerant electron, but defect-sensitive heat transport*.

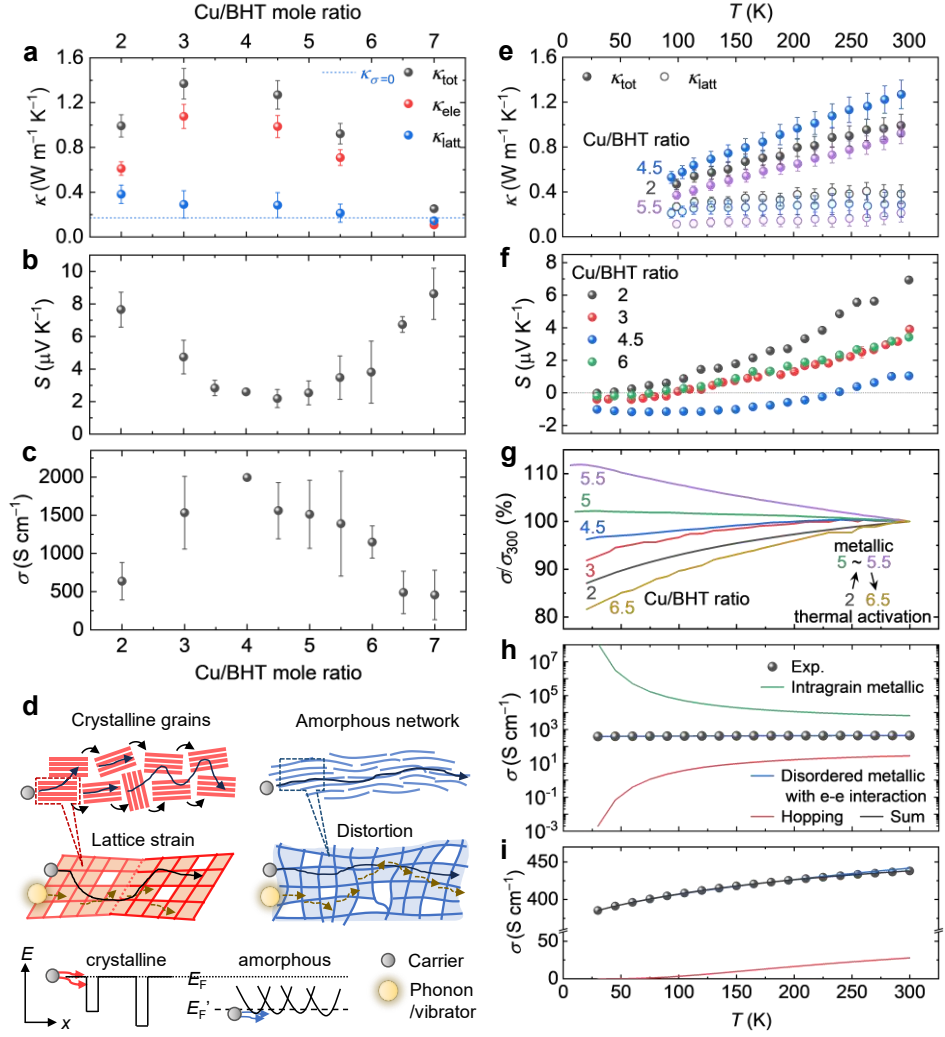

**Figure 3. Observation of defect-tolerant electron, but defect-sensitive heat transport in the temperature- and composition dependent thermoelectric coefficients. a – c**, Room temperature, in-plane thermal conductivities, Seebeck coefficients, and electrical conductivities of Cu-BHT films as a function of Cu/BHT ratio. In **a**, our experimental thermal conductivity at insulating state, i.e. lattice contribution at zero electrical conductivity, $\kappa_{\sigma=0}$ (blue dot line) is extracted from **Figure S15.** The error bars in **a, e** reflect the uncertainty associated with the standard deviation in the repeatability of measurements mainly due to variability in thermal and electrical contacts. The error bars in **b** and **c** represent sample standard deviation associated with experimental uncertainties, such as dimension determination and batch-to-batch variation in synthesis. The data reported here is averaged over 90 devices for conductivities and 57 devices for Seebeck coefficient across 14 different batches of films. Among them we only observed two devices showing negative Seebeck coefficient. **d**, Diagrams illustrating the interplay between chemical and structural (im)perfections, coherent charge carrier and incoherent lattice vibrations, and electronic landscape. **e, f**, Temperature-dependent in-plane thermal conductivities and Seebeck coefficients of different compositions. In **a** and **e**, the electronic and lattice contributions to the total thermal conductivities are evaluated from the Wiedemann-Franz law. **g,**

Temperature-dependent electrical conductivity as a function of Cu/BHT ratio where a transition from thermal activation to metallic transport is seen within a narrow range of compositions (ratio 5 – 5.5). **h, i**, Decomposition of the measured electrical conductivity of Cu/BHT ratio of 2 into contributions from intragrain metallic transport pathway and from disordered metallic and hopping transport pathways in the grain boundaries.

**Table 1. Summary of temperature coefficients of vibrational frequency, thermal expansion, and lattice thermal conductivities of different material systems.**

| Material system(s) | Temperature coefficient of vibrational frequency $|\chi|$ [a] ($10^{-2}$ cm$^{-1}$ K$^{-1}$) | Uniaxial thermal expansion coefficient $\alpha_{a,b,c}$ ($10^{-6}$ K$^{-1}$) | In-plane lattice thermal conductivity $\kappa$ (W m$^{-1}$ K$^{-1}$) |
|---|---|---|---|
| Organic molecules BTBT, DNTT, and their derivatives, Rubrene | 0.7 – 3.4 (diPh-BTBT and DNTT) [b]<br>0.9 – 10.1 (BTBT, diBu- and diC8-BTBT) | 16 – 36 ($\pi$-$\pi$ axis) [c]<br>30 – 120 (non-$\pi$-$\pi$ axis) [c] | 0.05 (C8DNTT)<br>0.25 (DNTT)<br>0.4 (Rubrene) |
| Organic polymers P3HT and PBTTT | – | – | 0.35 – 0.48 (P3HT)<br>0.39 (PBTTT) |
| Inorganic layered materials MoS$_2$, WS$_2$ | 1.2 | 1 – 8.6 | > 10 |
| Metal-organic coordination nanosheet Cu-BHT | 0.8 – 1.1 (S atoms displacement and C ring deformation, in-plane) [d] | 4.0 (in-plane) [e]<br>10.7 ± 0.80 (in-plane) [f]<br>38.3 ± 0.83 (cross-plane) [f] | 0.38 – 0.15 |

[a] Various vibration modes, including in-plane and out-of-plane. $\chi$ are negative for all materials here. [b] All forms of strong anharmonic behaviour suppressed. [c] Measured by X-ray diffraction. Rubrene not included. [d] See **Figure S13**, **Table S1** for details. [e] Evaluated by Raman spectroscopy with $\gamma$ found to be 0.87 for $N = 2$ (**Figure 4a**, **SI Section 3.2**). [f] Measured by X-ray diffraction (**Figures 4c–d**, **SI Section 3.1**).

## Characterisation of Vibrational Dynamics

In this section we focus on better understanding the lattice dynamics that underlies the defect-sensitive heat transport behavior of Cu-BHT films. To this aim, we performed temperature-dependent Raman spectroscopy and GIWAXS measurements on the most crystalline composition (Cu/BHT ratio of 2). Temperature-dependent Raman spectroscopy (Fig 4a) provides evidence for weak (perturbative) anharmonic effects in the lattice dynamics. As temperature increases from 4 to 300 K, the Raman bands are not sensitive to temperature (only shift of < 3 cm$^{-1}$ observed), with Grüneisen parameter $\gamma$ found to be 0.87 (for degeneracy of the Raman mode $N = 2$, see Supplementary Section 3.2 for details) within the quasi-harmonic approximation (QHA), comparable to that of inorganic solids and other 2D materials. The temperature coefficient of vibrational frequencies $\chi$ also show similar material-dependent tendency, as summarized in Table 1, Supplementary Table 1 and Fig. 13. Temperature-dependent grazing-incidence wide-angle X-ray scattering (GIWAXS) is able to unravel the anisotropic (in- vs cross-plane) lattice dynamics since thermal fluctuations can manifest themselves as changes in the intensity of Bragg peaks (non-cumulative disorder) or peak broadening (cumulative disorder, such as paracrystallinity)[20,40,41]. The uniaxial thermal expansion coefficients $\alpha$ are found to be $(10.7 \pm 0.80) \times 10^{-6}$ K$^{-1}$ for the in-plane direction and $(38.3 \pm 0.83) \times 10^{-6}$ K$^{-1}$ for the cross-plane direction (Supplementary Section 3.1), which is lower than organic semiconductors' in most cases and slightly larger than rigid inorganic and other 2D materials' (Table 1). It has been reported that in organic π-stacked materials low-frequency, large-amplitude vibrations, which are most harmful to charge delocalization, are suppressed in π-stacked direction with $\alpha = 16 - 36 \; 10^{-6}$ K$^{-1}$ because there is a sufficiently large force constant due to the presence of π–π interaction. Considering Cu-BHT's even shorter inter-layer distance compared to organic molecules' and its strong in-plane chemical bonds and expanded conjugated planes, the force constants acting against thermal displacements in all dimensions are expected to be larger than in van der Waals bonded organic semiconductors. Consistently both in-plane and cross-plane diffractions do not show apparent changes in intensity, line width, and line shape between 90 and 290 K (Fig. 4c,d, Supplementary Fig. 11), with pseudo-Voigt mixing parameter ($\eta$) remaining between 0.5 and 1 (Supplementary Fig. 11a,c) and only minimal shifts observed (Fig. 4c,d). These results suggest that in Cu-BHT, (1) large-amplitude vibrations are supressed and long-range ordering remains paracrystallinity-dominated (static disorder) upon thermal excitation and, (2) thermal fluctuations do not induce strong anharmonicity, pointing to a small-amplitude, quasi-harmonic lattice dynamics in all directions.

## Simulation of Vibrational Heat Carriers

To obtain a theoretical understanding of the exceptionally small lattice thermal conductivity, we performed first-principles simulations of heat conduction through lattice vibrations based on the Wigner Transport Equation (WTE)[38,39]. This framework generalizes the Boltzmann transport equation accounting for the vibrational heat not only carried by phonons that propagate particle-like, but also by wave-like tunneling between vibrational eigenstates with energy difference smaller than their energy uncertainty (linewidth). As a result, the WTE offers a comprehensive approach to predict the lattice thermal conductivity of a wide range of materials, including ordered-and-anharmonic crystals[39], disordered-and-harmonic glasses[42], as well as the intermediate regime of disordered-and-anharmonic 'complex crystals'[39].

For poor heat conductors such as Cu-BHT it is accurate to consider the WTE solution in the relaxation-time approximation[39,43]. Moreover, to compare with experiments in polycrystalline disordered samples where anisotropic transport could not be resolved, we consider the spatially averaged trace of the conductivity tensor,

$$\kappa(T) = \kappa_P(T) + \kappa_C(T) = \frac{1}{\mathcal{V}N_c} \sum_{\boldsymbol{q},s} \Bigg[ \tag{1}$$

$$C(\boldsymbol{q})_s \frac{\|\mathbf{v}(\boldsymbol{q})_{s,s}\|^2}{3} [\Gamma(\boldsymbol{q})_s]^{-1}$$

$$+ \sum_{s' \neq s} \frac{C(\boldsymbol{q})_s}{C(\boldsymbol{q})_s + C(\boldsymbol{q})_{s'}} \frac{\omega(\boldsymbol{q})_s + \omega(\boldsymbol{q})_{s'}}{2} \left( \frac{C(\boldsymbol{q}_s)}{\omega(\boldsymbol{q})_s} + \frac{C(\boldsymbol{q}_{s'})}{\omega(\boldsymbol{q})_{s'}} \right) \frac{\|\mathbf{v}(\boldsymbol{q})_{s,s'}\|^2}{3} \frac{\frac{1}{2}[\Gamma(\boldsymbol{q})_s + \Gamma(\boldsymbol{q})_{s'}]}{[\omega(\boldsymbol{q})_{s'} - \omega(\boldsymbol{q})_s]^2 + \frac{1}{4}[\Gamma(\boldsymbol{q})_s + \Gamma(\boldsymbol{q})_{s'}]^2} \Bigg].$$

Here, $C(\boldsymbol{q})_s = [\hbar^2 \omega^2(\boldsymbol{q})_s / k_b T^2] N(\boldsymbol{q})_s [N(\boldsymbol{q})_s + 1]$ is the specific heat at temperature $T$ of the vibration having wavevector $\boldsymbol{q}$, mode $s$, energy $\hbar\omega(\boldsymbol{q})_s$, and population given by Bose-Einstein distribution $N(\boldsymbol{q})_s = \left[\exp\left(\frac{\hbar\omega(\boldsymbol{q})_s}{k_B T}\right) - 1\right]^{-1}$; $\Gamma(\boldsymbol{q})_s$ is the total linewidth that accounts for anharmonic phonon-phonon scattering, electron-phonon scattering, presence of isotopic impurities, and scattering with grain boundaries (see SI for details). $\boldsymbol{v}(\boldsymbol{q})_{s,s'}$ is a velocity matrix—its diagonal elements $s = s'$ are the phonon group velocities, and its off-diagonal elements describe the strength of coherence's couplings between modes $s$ and $s'$ at the same $\boldsymbol{q}$; $N_c$ is the number of wavevectors sampling the Brillouin zone and $\mathcal{V}$ is the crystal's unit-cell volume. The total conductivity (1) accounts for both particle-like propagation and coherences' wave-like tunneling heat-transport mechanisms through $\kappa_P$ and $\kappa_C$, respectively. These macroscopic conductivities can be resolved in terms microscopic, single-vibration contributions, as shown by the two terms inside the square brackets: the first one describes phonons that carry heat $C(\boldsymbol{q})_s$ by propagating particle-like with velocity $\boldsymbol{v}(\boldsymbol{q})_{s,s}$ over lifetime $[\Gamma(\boldsymbol{q})_s]^{-1}$; the second one, instead, accounts for wave-like tunneling involving pairs of phonons $s, s'$ at the same wavector $\boldsymbol{q}$. It has been shown that in ordered and weakly anharmonic 'simple' crystals particle-like propagation dominates over wave-like tunneling ($\kappa_P \gg \kappa_C$)[38], in complex crystals both these mechanism co-exist ($\kappa_P \sim \kappa_C$), and in strongly disordered glasses around room temperature tunneling dominates ($\kappa_P \ll \kappa_C$)[44].

We based our simulations on the recently reported non-van der Waals layered structure of Cu$_3$BHT[17]. To justify the choice of this structure we analysed the non-van der Waals layered structure and the previously assumed 2D structure reported[13,14] from first principles and found that the non-van der Waals layered structure is energetically favoured[45] over the established 2D layered one (see Method section). Our predictions for the thermal conductivity (Fig. 4e) suggest that nearly pristine Cu-BHT (with micrometric or larger grains) features a total conductivity that decreases with temperature (crystal-like trend) between 100 and 300 K. This originates from having, in these ideal, ordered samples, particle-like transport mechanisms stronger than wave-like ones (Fig. 4f). We also note that at 300 K the conductivity of nearly pristine Cu-BHT is 1 W m$^{-1}$ K$^{-1}$ (Fig. 4e), a value that is still orders of magnitude lower than the conductivity of typical dense inorganic materials, highlighting the poor intrinsic lattice heat conduction in Cu-BHT. Such a property originates from the heavy copper atoms, which contribute to phonon modes with energy below 250 cm$^{-1}$ (Fig. 4b) and propagate heat poorly; this can be understood by noting that these modes feature low group velocity and relatively short (still

non-overdamped[39,43]) lifetime, conditions that imply a small contribution to the lattice thermal conductivity (see Eq. (1)). Accounting for structural disorder, with phonon scattering at the nanometric scale (using the Casimir model[46], see SI for details) yields dramatic changes in the macroscopic conductivity: (1) a strong reduction of the overall magnitude; (2) an inversion of the trend in temperature, from decreasing with $T$ (crystal-like) to increasing with $T$ (glass-like). Figure 4g demonstrates that these macroscopic changes result from the different effects of phonon-disorder scattering on microscopic heat transport mechanisms; in particular, while disorder suppresses particle-like propagation, it has a negligible impact on wave-like tunneling, making the latter the dominant conduction mechanism in the presence of strong disorder. Finally, we highlight how considering the theoretical minimum for the phonon scattering length due to disorder—the average bond length (~2 Å), also known as Kittel's limit[47]—yields predictions compatible with experiments (dotted-red line in Fig. 4e).

Our combined experimental and theoretical analyses therefore reveal that Cu-BHT films at and below room temperature feature: (1) quasi-harmonic lattice dynamics; and (2) disorder-driven, ultra-low temperature-activated lattice thermal conductivity. Specifically, structural disorder—likely a mixture of polycrystalline, paracrystalline, or amorphous domains—dominates over phonon-phonon and electron-phonon interactions in determining the conductivity. This is evidenced microscopically by the predominance of wave-like tunneling over particle-like propagation conduction mechanisms, leading to the thermal conductivity being strongly limited by the structural defects.

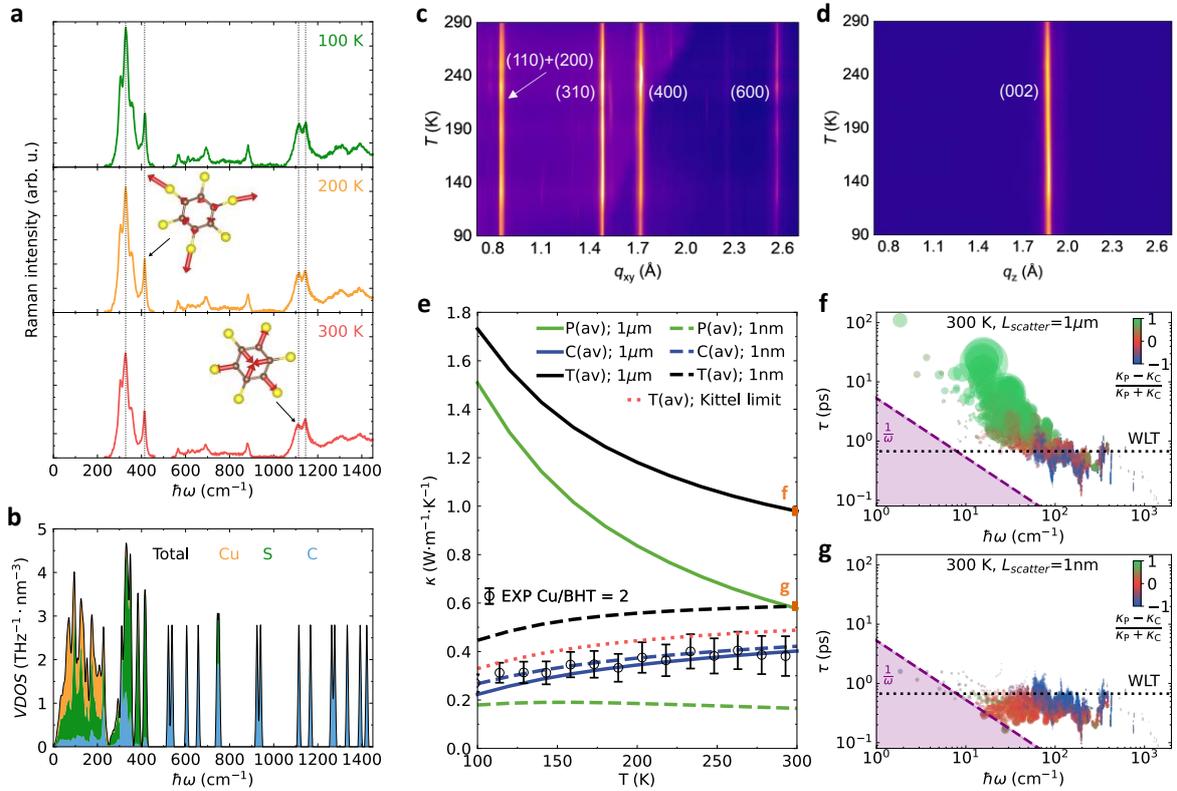

**Figure 4. Disorder-driven phonon-glass behavior in Cu-BHT**. **a**, Raman intensity as a function of temperature, with vertical lines showing negligible anharmonic shifts with temperature. The insets show the atomic displacement patterns of selected Raman-active modes, and panel **b** the first-principles predictions for Vibrational Density of States (VDOS) resolved into contributions from Cu, S and C atoms. **c,d** 2D maps of the temperature-dependent in-plane ($q_{xy}$) and out-of-plane ($q_z$) GIWAXS profiles, showing negligible peak shift and broadening in both directions. **e**, Thermal conductivity predictions as a function of temperature for various scattering lengths: 1 $\mu$m, 1nm, and the Kittel's theoretical limit (interatomic distance, ~2 Å). 'P' and 'C' indicate conductivity contributions from particle-like propagation and coherences' wave-like tunneling, respectively, and 'T' is their sum (total conductivity). Empty black circles are measurements in Cu/BHT ratio of 2 (same as in Fig. 3e). **f,g**, Impact of scattering length on total vibrations' lifetime and contribution to conductivity—the circles' sizes are proportional to conductivity contributions, and their color shows whether they behave particle-like (green), wave-like (blue) or a mix of the two (red is 50% of each). 'WLT' is the Wigner Limit in Time, a timescale determined by the material's structure that approximatively indicates at which lifetime vibrations transition from particle-like to wave-like behaviour[39].

## Nature of Charge Carriers

Low dynamic disorder in strongly electronically coupled materials with weak electron-phonon coupling promotes wave function delocalization and electron-crystal transport behavior, which can be probed through magnetotransport measurements. The magnetoresistance (MR), defined as [$\rho_{xx}$(B)–$\rho_{xx}$(0)]/$\rho_{xx}$(0), was found to be positive, exhibit a quadratic-field-dependence at all temperatures from 10-300 K in both crystalline Cu-BHT and films with paracrystallinity > 10% (Fig. 5b,c) and was much larger for a transverse orientation of the magnetic field with respect to the current direction than for a longitudinal orientation (Fig. 5a). The MR can be understood by carrier motion in a metallic regime or by wave function shrinkage in a hopping regime. The wave function shrinkage model for localised carriers in disordered materials is based on the magnetic field leading to a sharp decrease in the overlap of the wave functions on adjacent sites and a reduction in the hopping probability between sites[48,49]. However, a hopping model is not consistent with the weak temperature dependence of the conductivity. Therefore we interpret the MR as a signature of metallic transport; it can arise even in a unipolar transport regime due to anisotropies in the band structure, but is most commonly interpreted as a signature of an ambipolar transport regime, in which both electron and hole pockets on the Fermi surface contribute to transport, but exhibit different Hall angles[50]. The most crystalline sample (Cu/BHT ratio of 2) exhibits a lower transverse MR and weaker angular dependence at 14 T than the disordered, near amorphous sample (Cu/BHT ratio of 5; Fig. 5a–c, Supplementary Fig. 22). The difference is smaller than what would be expected from Kohler's rule, which states that the magnitude of the MR is determined by the magnitude of the magnetic field scaled by the zero field resistance $B/R$ (0,$T$). Kohler's rule is also found to be violated when analysing the temperature dependence of the MR (insets in Fig. 5b,c). This suggests that the observed differences in the transport properties of crystalline and amorphous samples as well as the temperature dependence of the MR do not merely reflect variations in carrier relaxation times, but more substantial changes in the transport physics.

To further elucidate this Hall effect measurements were performed. Interestingly, the temperature-dependent Hall coefficient $R_H$ (Fig. 5d, Supplementary Fig. 23) is positive, i.e. hole dominated, at all temperatures and exhibits a peak at around 135 K for the metallic composition with paracrystallinity > 10%. In contrast, in the most crystalline films we observed a sign change of the Hall coefficient around 210 K with transport at higher temperatures becoming electron dominated. We note that the Hall results here might not be directly quantitatively comparable to the results of Seebeck coefficient, which shows a hole-dominated behaviour near room temperature. The Seebeck and Hall/MR measurements were not performed on the same samples and the device preparation and loading of the devices for magnetotransport measurements required some unavoidable aging upon light and air exposure, which could have caused a shift of the Fermi energy within the electronic structure. However, it is still clear that there is a sign change observed in the Seebeck coefficient as a function of temperature as well. These observations therefore strongly suggest that a two-band transport model with contributions from both electron and hole pockets within a metallic electronic structure[51,52] does indeed need to be considered to consistently explain the MR and Hall data (Fig. 5e).

In the disordered composition the UPS measurements indicate a comparatively high workfunction (Supplementary Fig. 9), possibly due to p-type doping by the defects that are present. The Fermi level is presumably at a position within the hole pocket but near to the bottom of the electron pocket, so that below the peak in Hall coefficient at 135 K transport is hole dominated, but above a thermally generated population of electrons is beginning to occupy the electronic states of the electron pocket and contribute to the transport. In the crystalline samples the workfunction is reduced, i.e. the Fermi level shifts up to a position where it crosses both bands. In this regime a hole-electron competing transport is at play. This is supported by our simulated electronic band structure of Cu-BHT, which shows both electron and hole pockets contributing at the Fermi level.

Based on these results we developed a two-carrier transport model that is consistent with the temperature dependent conductivity, MR and Hall data, and allows us to estimate hole and electron carrier concentrations $n_h$, $n_e$ and mobilities $\mu_h$, $\mu_e$. In principle, this is an underdetermined problem and we need to make certain approximations in the model. We assume, in particular, that the ratio of hole and electron mobility $b$ is a temperature-independent constant (see Supplementary Section 6 together with details of the simulations). The simulated results show some clear and consistent trends, that is, $\mu_e > \mu_h$ and $n_e < n_h$, for both amorphous and crystalline samples:

(i) Sample with paracrystallinity > 10%: $\mu_e = 17 - 37$ cm$^2$ V$^{-1}$ s$^{-1}$; $\mu_h = 11 - 17$ cm$^2$ V$^{-1}$ s$^{-1}$; $n_e = 0.42 - 2.6 \times 10^{20}$ cm$^{-3}$; $n_h = 5.1 - 9.1 \times 10^{20}$ cm$^{-3}$;

(ii) Crystalline composition: $\mu_e = 6.4 - 13.8$ cm$^2$ V$^{-1}$ s$^{-1}$; $\mu_h = 5.2 - 6.4$ cm$^2$ V$^{-1}$ s$^{-1}$; $n_e = 0.26 - 1.07 \times 10^{21}$; $n_h = 0.82 - 1.5 \times 10^{21}$ cm$^{-3}$

The ranges indicated reflect uncertainties in the approximations made in the model. For comparison, $2.1 \times 10^{21}$ cm$^{-3}$ corresponds to 1 carrier per unit cell. We attribute the higher mobilities of disordered, near amorphous samples to a smoother energetic landscape that less significantly scatters holes and traps electrons. The observation of weakly increasing $\mu_e$ and $\mu_h$ with decreasing temperature in both crystalline and near amorphous samples consistently suggest that carrier motions in Cu-BHT are mainly in a (disordered) metallic regime, that is *defect-tolerant*.

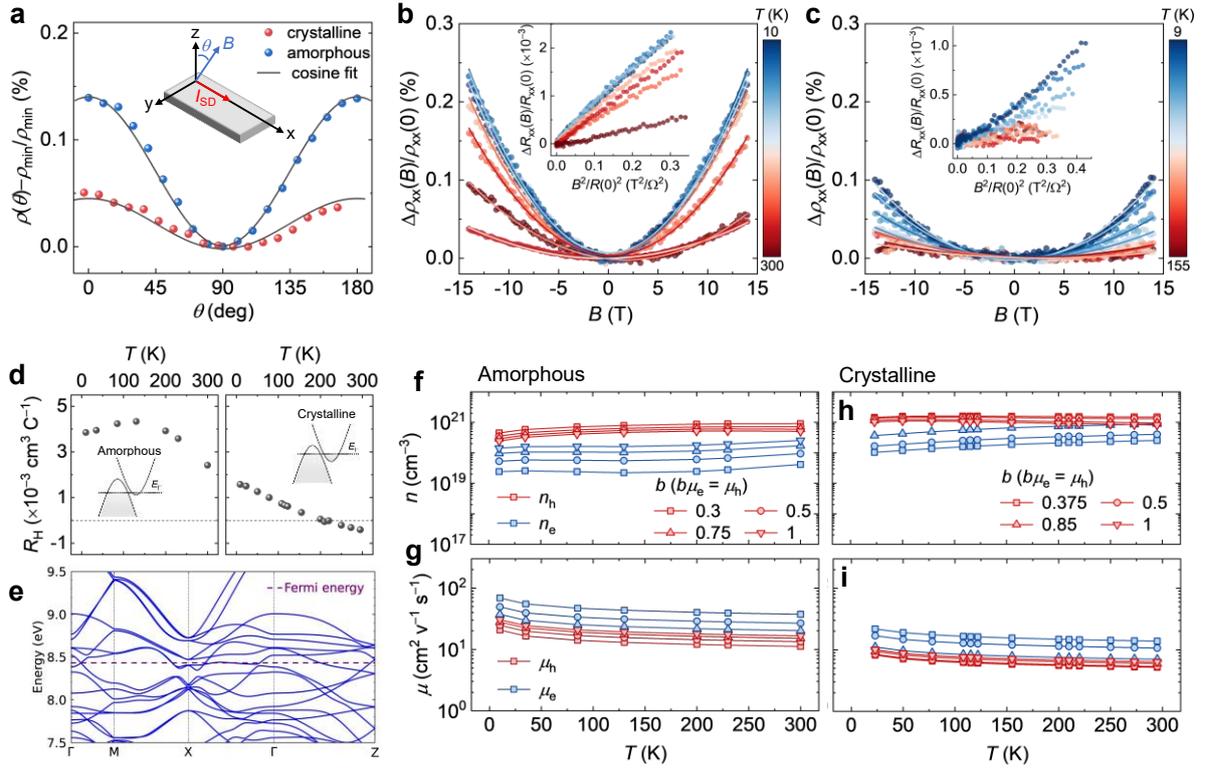

**Figure 5. Defect-manipulated anisotropic, electron-crystal magnetotransport in crystalline and near amorphous Cu-BHT films. a**, Angular-dependent magnetoresistance measurements at 14 T and 50 K showing minima at $\theta \sim 90$ for both amorphous Cu/BHT ratio 5 and crystalline ratio 2. The architecture of the measurements is illustrated in the inset. **b, c**, Quadratic magnetoresistance for amorphous and crystalline samples respectively ($\theta = 0$). Corresponding Kohler's plots are shown in the inset. **d**, Temperature-dependent Hall coefficient of amorphous (left) and crystalline (right) compositions, with inset diagrams illustrating different Fermi levels at two-band transport region. **e**, Electronic band structure. **f, g**, Simulated temperature-dependent carrier concentrations for hole and electron and their mobilities of the amorphous sample. **h** and **i**, Extracted temperature-dependent carrier concentrations for hole and electron and their mobilities of the crystalline sample. $b$ is a scaling factor, $b\mu_e = \mu_h$.

## Conclusions

We have investigated the structure-property relationships and thermoelectric transport physics of Cu-BHT, as a model system for conjugated coordination nanosheets. We have found the charge transport to be *defect-tolerant*—metallic electrical conductivities > 1000 S/cm can be reached in films that contain a high density of structural defects and exhibit paracrystallinity values > 10%. This defect tolerance is unprecedented and surprising, in sharp contrast with traditional silicon or molecular (semi)conductors in which electron transport is more strongly thermally activated in more disordered, amorphous systems. Importantly, this *defect-tolerant electron transport* in Cu-BHT coexists with *defect-sensitive heat transport* with a lattice thermal conductivity that is exceptionally low and practically at the theoretical limit. We have shown that this originates from having low-amplitude lattice vibrations, which are quasi-harmonic, but for which the associated thermal conductivity is predominantly wave-like and strongly limited by defects and damped by structural disorder, in sharp contrast with the *defect-tolerant* behaviour observed for electrons. This thermoelectric transport regime in disordered Cu-BHT is analogous to the electron-crystal, phonon-glass behaviour that is observed in some highly crystalline inorganic thermoelectric materials. On the basis of Hall effect and magnetoresistance measurements as well as first-principles calculations we have argued that Cu-BHT exhibits a hole-electron competing transport regime. This is likely to reduce the Seebeck coefficient and currently does not allow reaching competitive thermoelectric performance levels. For thermoelectric applications the insufficient Seebeck coefficient is the key challenge that remains to be overcome. If it was possible to tune the Fermi level into an effective, one-band regime and control the unipolar carrier concentration, conjugated coordination nanosheets could be very promising materials for thermoelectric as well as (bio)electronic applications.

## Methods

**Synthesis and film transfer**

Adapted from literatures[13,16], we improve the liquid-liquid interfacial synthesis of Copper benzenehexathiol (Cu-BHT) in this work. The organic solvent used here is a mixture of chloroform (CF) and chlorobenzene (CB) in a ratio of CF:CB = 4.5:1, v/v. Under nitrogen atmosphere, BHT is first dissolved in degassed organic solvent at room temperature to provide a solution in a concentration of 1.11 mM. After filtered by a 0.45 um PTFE membrane, BHT solution of 1 mL is added into a small vial (20 mL volume) containing mixed organic solvent of 4.5 mL. Filtered degassed water of 6.5 mL is then added onto the organic phase to form an interface. Depending on the desired Cu/BHT ratio, certain amount of aqueous solution of $Cu(CH_3COO)_2$ in a concentration of 5 mM is filtered and added onto the water at 45 °C very gently (not to fluctuate the oil-water interface). Lastly the reaction system is sealed and standing at 45 °C for 2 hours in nitrogen.

To transfer the Cu-BHT films onto substrates for characterizations, first, the upper layer, aqueous solution, is partly removed and diluted, washed with water very gently without fluctuating the interface, in order to remove residual $Cu^{2+}$ and water-soluble impurities. Then an $O_2$ plasma-cleaned, facing down substrate passes through the interface; at this point, the Cu-BHT film attaches on it. The substrate with the film is held on in the organic phase, simultaneously the aqueous layer is fully removed to expose the surface of the organic solution. Finally, the substrate with a film on it is slowly taken out. After naturally drying out, the substrate with film attached can be immersed into organic solvents, ethanol, followed by CF, for washing without lift off (except for methanol, which leads to lift off). These specific procedures usually generate films with thickness ranging from 100 – 400 nm, depending on the amount of $Cu^{2+}$ added.

Note that the purity of the starting material BHT is significant for the electronic properties of the Cu-BHT films. Light yellow BHT that contains impurities impedes electronic conduction to enter metallic regime (see Supporting Information section 5.4 for details). All samples investigated in this work follow above procedures unless otherwise specified.

**Scanning electron microscopy (SEM) and energy dispersive X-ray spectroscopy (EDX)**

The surface morphologies of the Cu-BHT films were imaged using a Zeiss LEO 1550 field-emission SEM with a working distance of 3.5 – 4 mm, an acceleration voltage of 3 kV, and an in-lens detector. The element composition was acquired using a dual beam microscope Helios FIB-SEM equipped with Oxford Instruments EDS detector at the cross section of the films. An acceleration voltage of 20 kV, working distance of 4 mm, and an acquisition time of 3 – 5 minutes were used during EDX measurement where no drift was seen. Samples for these two measurements were deposited on Si without thermal-growth oxide.

**Grazing-incidence wide-angle X-ray scattering (GIWAXS)**

GIWAXS measurements were performed at Diamond Light Source beamline I-07 at 12.5 keV X-ray beam energy. Samples were prepared on 1 cm x 1 cm Si wafer without thermal-growth oxide. Images were collected using a Pilatus 2m camera positioned 450.5 mm from the sample. Sample-detector distance was calibrated using a silver behenate reference sample. All measurements were performed at an incidence angle of 0.2 degree, with samples mounted on a temperature-controlled stage inside a helium filled chamber with Kapton windows. Temperature-dependent measurements consisted of two 5 second exposures at attenuation level 2 (9.9% of full beam intensity) with shifted camera position to gap-fill detector gap regions. A full sample realignment was performed between temperature steps to ensure consistent sample-detector distance, which could otherwise lead to systematic errors in the measured reciprocal spacing $q$. Data was processed using the MATLAB package GIXSGUI[53].

**Scanning electron diffraction (SED)**

To fabricate thin, electron-transparent Cu-BHT films, the reaction time was shortened to 15 – 20 minutes, yielding films of 40 – 70 nm. The resulting thin films were transferred onto O$_2$-plasma-cleaned SiN$_x$ grids with a 30 nm-thick, low-stress amorphous Si$_3$N$_4$ membrane window (NT025X, Norcada). Care was taken during transferring and optical microscope and SEM were employed to check the samples. The procedures of acquiring low-dose SED microscopy and the data processing are reported in previous work[54]. During acquisition, a two-dimensional (2D) electron diffraction pattern was measured at every probe position of a scanning electron beam in transmission mode. SED data was acquired on the JEOL ARM300CF E02 instrument at ePSIC (Diamond Light Source, Didcot-Oxford, UK). An acceleration voltage of 200 keV, nanobeam alignment (convergence semiangle) of 1 mrad, electron probe of 5 nm, a scan dwell time of 1 ms, a fluence of ~11 e$^-$Å$^{-2}$, and camera length of 20 cm were used. Post-processing of SED data was done using pyXem 0.14[55] (an open-source Python library for crystallographic electron microscopy). All diffraction patterns were distortion-corrected and calibrated with an Au cross grating. The drift in the beam position of the non-scattered beam was corrected and centered for all frames using cross-correlation with a subpixel factor of 10. To display diffraction planes that match real-space features, all diffraction patterns shown in this work were rotation-corrected using a MoO$_3$ calibration sample. Dead pixels and detector junctions were masked. The analysis of the processed SED data was done using the Automated Crystal Orientation Mapping in py4DSTEM 0.13.6[56]. To obtain the orientation map, peak finding through template matching was run on each probe position and was compared to a simulated library (from the CIF file in Supplementary Information) using sparse correlation matching. The flow map was generated by radially binning diffraction patterns in the range of 0.1 to 1.25 Å$^{-1}$ and obtaining the strongest angular direction. Virtual dark field images were generated from selecting specific diffraction spits and mapping its intensity as a function of probe position.

**Raman spectroscopy**

Raman spectra were collected by using a Horiba T64000 Raman spectrometer under a laser excitation of 532 nm. Samples were prepared on either 1 cm x 1 cm Si wafer without thermal-growth oxide or Corning EAGLE XG glass. Before acquisition, spectrometer was calibrated with the characterized band (520.70 cm$^{-1}$) of a standard Si sample. Regular measurements were performed in ambient air at room temperature and temperature-dependent measurements were under high vacuum with liquid Helium as cooling system. During acquisition, integration time of no less than 30 s, 10 cycles, and a laser power that did not lead to visible beam damage were used.

**Ultraviolet photoelectron spectroscopy (UPS)**

UPS experiments for measuring valence bands and work functions were carried out on the films transferred onto n-doped Si without thermal-growth oxide by using an ultrahigh-vacuum photoemission instrument, Escalab 250Xi, with a 21.22 eV excitation source. The samples had not been exposed to air by using a transfer tube for sample transferring and loading. During scanning an electrical bias of -5 V was applied to the samples. Scanning steps of 0.05 and 0.01 eV were respectively used for full-range spectra showing the secondary electron cutoff and, for the narrow, near zero binding energy range which shows the edge of the valence bands.

**Device fabrication**

For initial electrical conductivity and Seebeck coefficient measurements, a four-parallel-electrode device architecture was used. For all temperature-dependent measurements – electrical conductivity, Seebeck coefficient, DC and AC Hall effects, and magnetotransport – a multifunction device architecture was employed. Substrates (Corning EAGLE XG, thickness ~ 700 $\mu$m) were cleaned by sequential sonication steps in acetone, 2% Decon 90 / DI water, DI water, and isopropanol (10 minutes for each). Then washed substrates were dried with nitrogen gas and exposed to oxygen plasma at 300 W for 3 – 5 minutes. All electrical contacts, Cr/Au (4 nm/20 nm), were deposited on the freshly prepared, cleaned substrates by thermal evaporation through shadow mask method for four-parallel-electrode devices and UV lithography for multifunction devices. Cu-BHT films were transferred onto them as Section 1.1 described.

**Room-temperature electrical conductivity and Seebeck coefficient measurements**

Measurements were performed on a manual probe station using an Agilent 4155B Semiconductor Parameter Analyzer for electrical conductivity measurement and using a Keithley Nanovoltmeter 2182A for Seebeck coefficient measurement under nitrogen atmosphere (Belle Technology, < 5 ppm $O_2$ and < 15 ppm $H_2O$). Each device was electrically isolated before measurement by carefully scratching off the film outside the active device area under an optical microscope. Film thickness was measured by surface profilimetry.

**Temperature-dependent electrical conductivity and Seebeck coefficient measurements**

For temperature-dependent thermoelectric measurements, data were recorded with multifunction devices in a LakeShore Cryotronics CRX-4K probe station equipped with closed-cycle liquid Helium under high vacuum ($10^{-7}$ to $10^{-6}$ mbar). A Keithley 2182A nanovoltmeter (for thermovoltage determination) and a couple of Keithley 2612B source-meters (for voltage sourcing and four probe conductivity measurements) were used. The methodology of the Seebeck coefficient measurement is adapted from previous work by Venkateshvaran *et al.*[57] and by Statz *et al.*[58] Error analysis was performed following the report by Statz *et al.*[58]

**Room-temperature and temperature-dependent thermal conductivity measurements**

Cu-BHT films were deposited onto specially designed, commercially available silicon-based chips containing two free standing, different-area $Si_3N_4$ membranes (refer to as Linseis chips). Two microfabricated electrical wires aligned with the longitudinal axes of the membranes serve as heater and thermometer. A high width/thickness ratio of the membranes on which Cu-BHT films covered ensures that the heat flux is predominantly horizontally one-dimensional across the sample and in line with the plane of the membranes. Thus, the measurements performed in this work probe the in-plane thermal conductivities of the samples by using a $3\omega$-based method. By performing measurements on the two different-area membranes integrated on the same substrate, correction for radiative losses and subtraction of the contribution from empty membranes to total heat conduction (from membrane and sample) is allowed; accurate measurements are hence enabled. Calculation of thermal conductivity was carried out on the raw data in the software of Linseis Thin-Film Analyzer according to the method developed by Linseis *et al.*[59]

**First-principles predictions for the vibrational and thermal properties**

Since past works discussed the layered Cu-BHT structure to naturally form in the AA stacking pattern, our simulations started comparing the energy of the different known structures for AA stacked layered Cu-BHT: (1) the established AA stacked structure, which contains 15 atoms per primitive cell[13,14,16], and (2) the recently observed 60-atom structure[17]. Starting from the experimental primitive cells[17], we relaxed atomic positions and lattice vectors using Density Functional Theory (DFT) as implemented in Quantum Espresso[60,61]. We employed the PBEsol functionals with Grimme-D2 van der Waals corrections, kinetic energy cutoff for charge density and wavefunctions equal to 720 Ry and 90 Ry, respectively, $6.6(6) \times 10^{-7}$ Ry/atom energy threshold, and $6.6(6) \times 10^{-7}$ Ry/Bohr/atom force threshold, and 0.05 kbar pressure convergence threshold. A Marzari-Vanderbilt smearing equal to 0.02 Rydberg was used. The Brillouin zones were sampled using a Monkhorst-Pack meshes equal to 4x4x8 and 2x4x4 for 15 and 60 atom structures respectively with a zero shift for both structures. We employed pseudopotentials taken from the SSSP efficiency library[62].

After complete relaxation, the energy per atom of the 15-atom cell (-97.5625 Ry/atom) resulted higher than that of the 60-atom cell (-97.5634 Ry/atom), indicating that the latter is energetically favoured at this level of theory. We have also verified that performing calculations at the PBEsol + Grimme-D3 level yields consistent results (-97.5590 Ry/atom and -97.5599 Ry/atom for 15 and 60 atom structures respectively). Given that the 60-atom cell was found to be energetically favored[45], we focused on it in the remainder of the analysis. In addition, given the unimportant differences between the PBEsol+D2 and PBEsol+D3 level of theory, and accounting for the fact that in Quantum Espresso Density Functional Perturbation Theory (DFPT, needed to compute the electron-phonon contribution to the linewidth $\Gamma_s(\boldsymbol{q})$ appearing in Eq. (1)) is implemented at the PBEsol+D2 but not PBEsol+D3 level, we proceeded using DFPT with PBEsol+D2.

Second and third-order force constants were obtained using hiphive[63] in the 60-atom cell with 120 rattled structures; we used cutoffs equal to the default value 3.0 Angstrom for both second- and third-order force constants. Electron-phonon couplings were calculated using DFPT, and given the large size of the primitive cell of non van der Waals layered Cu-BHT (60 atoms), a Gamma-point (**q**=**0** only) calculation was performed. The Wigner conductivity expression (1) was evaluated accounting for electron-phonon[64,65], phonon-phonon[66,67], phonon-isotope[68], and phonon disorder scattering[46] (see SI for details). Electron-phonon linewidths on a dense mesh were obtained using Wannier-interpolation, as implemented in wannier90[69,70] and Phoebe[65], using a computationally converged 5x9x13 q-mesh for phonons and 25x45x65 k-mesh for electrons. Bulk lattice thermal conductivity calculations were performed using Phoebe on a 5x9x13 q-mesh (See Supplementary Fig. 15), accounting for phonon-phonon, phonon-isotope and electron-phonon scattering. Phonon-disorder scattering was accounted for by postprocessing the Phoebe output linewidth. The Dirac delta appearing in the linewidth expression (see SI section 4.2) was broadened with a Gaussian smearing equal to $4.92 \times 10^{-5}$ eV.

**Temperature-dependent DC Hall effect and magnetoresistance measurements**
DC Hall effect and magnetoresistance measurements were carried out on Quantum Design 14T-PPMS DynaCool D-134 with either standard or rotator sample pucks. Multifunctional Hall bar device architecture was used and electrical connection between the device and the sample puck was built by wire bonding. Four-point-probe electrical conductivity at zero field was also recorded as a function of temperature during the measurement.

## Data availability

All the data supporting the findings of this study are included in the Article and its Supplementary Information files, and are available from the corresponding author on reasonable request.

## References

1. Moore, A. R. Electron and hole drift mobility in amorphous silicon. *Appl. Phys. Lett.* **31**, 762–764 (1977).
2. Tian, H. *et al.* Disorder-tuned conductivity in amorphous monolayer carbon. *Nature* **615**, 56–61 (2023).
3. Mu, X., Wu, X., Zhang, T., Go, D. B. & Luo, T. Thermal transport in graphene oxide - From ballistic extreme to amorphous limit. *Sci. Rep.* **4**, 1–9 (2014).
4. Das, P., Bathula, S. & Gollapudi, S. Evaluating the effect of grain size distribution on thermal conductivity of thermoelectric materials. *Nano Express* **1**, 020036, (2020).
5. Berland, K. *et al.* Enhancement of thermoelectric properties by energy filtering: Theoretical potential and experimental reality in nanostructured ZnSb. *J. Appl. Phys.* **119**, (2016).

6. Maeda, H., Takada, K., Fukui, N., Nagashima, S. & Nishihara, H. Conductive coordination nanosheets: Sailing to electronics, energy storage, and catalysis. *Coord. Chem. Rev.* **470**, 214693 (2022).

7. Sakamoto, R. *et al.* Layered metal-organic frameworks and metal-organic nanosheets as functional materials. *Coord. Chem. Rev.* **472**, 214787 (2022).

8. Wang, M., Dong, R. & Feng, X. Two-dimensional conjugated metal-organic frameworks (2Dc-MOFs): chemistry and function for MOFtronics. *Chem. Soc. Rev.* **50**, 2764–2793 (2021).

9. Yong, X. *et al.* Tuning the thermoelectric performance of π-d conjugated nickel coordination polymers through metal-ligand frontier molecular orbital alignment. *J. Mater. Chem. A* **6**, 19757–19766 (2018).

10. Deng, T. *et al.* 2D Single-Layer π-Conjugated Nickel Bis(dithiolene) Complex: A Good-Electron-Poor-Phonon Thermoelectric Material. *Adv. Electron. Mater.* **5**, 1800892, (2019).

11. He, Y. *et al.* Two-dimensional metal-organic frameworks with high thermoelectric efficiency through metal ion selection. *Phys. Chem. Chem. Phys.* **19**, 19461–19467 (2017).

12. Un, H. I. *et al.* Controlling Film Formation and Host-Guest Interactions to Enhance the Thermoelectric Properties of Nickel-Nitrogen-Based 2D Conjugated Coordination Polymers. *Adv. Mater.* **36**, 1–11 (2024).

13. Huang, X. *et al.* Superconductivity in a Copper(II)-Based Coordination Polymer with Perfect Kagome Structure. *Angew. Chemie* **130**, 152–156 (2018).

14. Huang, X. *et al.* A two-dimensional π-d conjugated coordination polymer with extremely high electrical conductivity and ambipolar transport behaviour. *Nat. Commun.* **6**, 7408, (2015).

15. Tsuchikawa, R. *et al.* Unique Thermoelectric Properties Induced by Intrinsic Nanostructuring in a Polycrystalline Thin-Film Two-Dimensional Metal–Organic Framework, Copper Benzenehexathiol. *Phys. Status Solidi Appl. Mater. Sci.* **217**, 2000437 (2020).

16. Toyoda, R. *et al.* Heterometallic Benzenehexathiolato Coordination Nanosheets: Periodic Structure Improves Crystallinity and Electrical Conductivity. *Adv. Mater.* **34**, 2106204 (2022).

17. Pan, Z. *et al.* Atomic-Precision Non-van der Waals 2D Structures: Superconductivity in π-d Conjugated Coordination Polymers. *ChemRxiv* DOI: 10.26434/chemrxiv-2024-296c0 (2024).

18. Wu, Y. *et al.* Lattice Strain Advances Thermoelectrics. *Joule* **3**, 1276–1288 (2019).

19. Hanus, R. *et al.* Lattice Softening Significantly Reduces Thermal Conductivity and Leads to High Thermoelectric Efficiency. *Adv. Mater.* **31**, 1900109, (2019).

20. Rivnay, J., Noriega, R., Kline, R. J., Salleo, A. & Toney, M. F. Quantitative analysis of lattice disorder and crystallite size in organic semiconductor thin films. *Phys. Rev. B - Condens. Matter Mater. Phys.* **84**, 1–20 (2011).

21. Prosa, T. J., Moulton, J., Heeger, A. J. & Winokur, M. J. Diffraction line-shape analysis of poly(3-dodecylthiophene): a study of layer disorder through the liquid crystalline polymer transition. *Macromolecules* **32**, 4000–4009 (1999).

22. Kambe, T. *et al.* π-Conjugated nickel bis(dithiolene) complex nanosheet. *J. Am. Chem. Soc.* **135**, 2462–2465 (2013).

23. Park, J. *et al.* Synthetic Routes for a 2D Semiconductive Copper Hexahydroxybenzene Metal-Organic Framework. *J. Am. Chem. Soc.* **140**, 14533–14537 (2018).

24. Dou, J. H. *et al.* Atomically precise single-crystal structures of electrically conducting 2D metal–organic frameworks. *Nat. Mater.* **20**, 222–228 (2021).

25. Van Lien, N. & Dinh Toi, D. Coulomb correlation effects in variable-range hopping thermopower. *Physics Letters A* **261**, 108 (1999).

26. Zabrodskii, A. G. & Zinov'eva, K. N. Low-Temperature Conductivity and Metal-Insulator Transition in Compensate n-Ge. *Sov. Phys. JETP* **59**, 425–433 (1984).

27. Mott, S. N. The mobility edge since 1967. **20**, 657–684 (1995)

28. Mott, N. F. The conductivity near a mobility edge. *Philos. Mag. B Phys. Condens. Matter; Stat. Mech. Electron. Opt. Magn. Prop.* **49**, 75–82 (1984).

29. Xie, J. *et al.* Intrinsic glassy-metallic transport in an amorphous coordination polymer. **611**, 479, *Nature* (2022)

30. Ito, H., Mada, H., Watanabe, K., Tanaka, H. & Takenobu, T. Charge transport and thermoelectric conversion in solution-processed semicrystalline polymer films under electrochemical doping. *Commun. Phys.* **4**, 8, (2021).

31. Tang, H. *et al.* A solution-processed n-type conducting polymer with ultrahigh conductivity. *Nature* **611**, 271–277 (2022).

32. Kaiser, A. B. Systematic conductivity behavior in conducting polymers: Effects of heterogeneous disorder. *Adv. Mater.* **13**, 927–941 (2001).

33. Tran, V. A., Tran, P. A., Nguyen, H. Q., Bach, G. H. & Nguyen, T. T. Boundary-scattering induced Seebeck coefficient enhancement in thin films within relaxation time approximation. *Phys. B Condens. Matter* **635**, 413800, (2022).

34. Sahoo, S., Gaur, A. P. S., Ahmadi, M., Guinel, M. J. F. & Katiyar, R. S. Temperature-dependent Raman studies and thermal conductivity of few-layer $MoS_2$. *J. Phys. Chem. C* **117**, 9042–9047 (2013).

35. Ushirokita, H. & Tada, H. In-plane thermal conductivity measurement of conjugated polymer films by membrane-based AC calorimetry. *Chem. Lett.* **45**, 735–737 (2016).

36. Okada, Y. *et al.* Low-temperature thermal conductivity of bulk and film-like rubrene single crystals. *Phys. Rev. B - Condens. Matter Mater. Phys.* **83**, 3–6 (2011).

37. Selezneva, E. *et al.* Strong Suppression of Thermal Conductivity in the Presence of Long Terminal Alkyl Chains in Low-Disorder Molecular Semiconductors. *Adv. Mater.* **33**, 2008708, (2021).

38. Simoncelli, M., Marzari, N. & Mauri, F. Unified theory of thermal transport in crystals and glasses. *Nat. Phys.* **15**, 809, (2019)

39. Simoncelli, M., Marzari, N. & Mauri, F. Wigner Formulation of Thermal Transport in Solids. *Phys. Rev. X* **12**, 041011, (2022)

40. Troisi, A. & Cheung, D. L. Transition from dynamic to static disorder in one-dimensional organic semiconductors. *J. Chem. Phys.* **131**, 014703, (2009).

41. Zhang, W. *et al.* Thermal Fluctuations Lead to Cumulative Disorder and Enhance Charge Transport in Conjugated Polymers. *Macromol. Rapid Commun.* **40**, 1–11 (2019).

42. Simoncelli, M., Mauri, F. & Marzari, N. Thermal conductivity of glasses: first-principles theory and applications. *npj Comput. Mater.* **9**, 106, (2023)

43. Pazhedath, A., Bastonero, L., Marzari, N. & Simoncelli, M. First-principles characterization of thermal conductivity in LaPO4-based alloys. *Phys. Rev. Appl.* **10**, 1 (2023).

44. Harper, A. F., Iwanowski, K., Witt, W. C., Payne, M. C. & Simoncelli, M. Vibrational and thermal properties of amorphous alumina from first principles. *Phys. Rev. Mater.* **8**, 1–19 (2024).

45. Pickard, C. J. & Needs, R. J. Ab initio random structure searching. *Journal of Physics Condensed Matter* **23**, 053201, (2011).

46. Fugallo, G., Lazzeri, M., Paulatto, L. & Mauri, F. Ab initio variational approach for evaluating lattice thermal conductivity. *Phys. Rev. B - Condens. Matter Mater. Phys.* **88**, 045430, (2013)

47. Kittel, C. Interpretation of the thermal conductivity of glasses. *Phys. Rev.* **75**, 972, (1949).

48. Gu, H. *et al.* Electrical transport and magnetoresistance in advanced polyaniline nanostructures and nanocomposites. *Polymer (Guildf).* **55**, 4405–4419 (2014).

49. Huang, S. M., Yu, S. H. & Chou, M. Two-carrier transport-induced extremely large magnetoresistance in high mobility Sb2Se3. *J. Appl. Phys.* **121**, 015107, (2017).

50. Li, Y. *et al.* Hole pocket–driven superconductivity and its universal features in the electron-doped cuprates. *Sci. Adv.* **5**, 1–8 (2019).

51. LeBoeuf, D. *et al.* Electron pockets in the Fermi surface of hole-doped high-Tc superconductors. *Nature* **450**, 533–536 (2007).

52. Androulakis, J. *et al.* Thermoelectric enhancement in PbTe with K or Na codoping from tuning the interaction of the light- and heavy-hole valence bands. *Phys. Rev. B - Condens. Matter Mater. Phys.* **82**, 1–8 (2010).

53. Jiang, Z. GIXSGUI: A MATLAB toolbox for grazing-incidence X-ray scattering data visualization and reduction, and indexing of buried three-dimensional periodic nanostructured films. *J. Appl. Crystallogr.* **48**, 917–926 (2015).

54. Ferrer Orri, J. *et al.* Unveiling the Interaction Mechanisms of Electron and X-ray Radiation with Halide Perovskite Semiconductors using Scanning Nanoprobe Diffraction. *Adv. Mater.* **34**, 2200383 (2022).

55. D. N. Johnstone, P. Crout, S. Høgås, B. Martineau, J. Laulainen, S. Smeets, S. Collins, E. Jacobsen, J. Morzy, E. Prestat, T. Doherty, T. Ostasevicius, H. W. Ånes, T. Bergh, R. Tovey,

E. O. pyxem/pyxem: pyxem 0.12.0. https://doi.org/10.5281/zenodo.3968871.

56. Ophus, C. *et al.* Automated Crystal Orientation Mapping in py4DSTEM using Sparse Correlation Matching. *Microsc. Microanal.* **28**, 390–403 (2022).

57. Venkateshvaran, D. *et al.* Approaching disorder-free transport in high-mobility conjugated polymers. *Nature* **515**, 384–388 (2014).

58. Statz, M. *et al.* On the manifestation of electron-electron interactions in the thermoelectric response of semicrystalline conjugated polymers with low energetic disorder. *Commun. Phys.* **1**, 16, (2018).

59. Linseis, V., Völklein, F., Reith, H., Nielsch, K. & Woias, P. Advanced platform for the in-plane ZT measurement of thin films. *Rev. Sci. Instrum.* **89**, 015110, (2018).

60. Paolo, G. *et al.* QUANTUM ESPRESSO: a modular and open-source software project for quantum simulations of materials. *J. Phys. Condens. Matter* **21**, 395502, (2009).

61. Giannozzi, P. *et al.* Advanced capabilities for materials modelling with Quantum ESPRESSO. *J. Phys. Condens. Matter* **29**, 465901, (2017)

62. Prandini, G., Marrazzo, A., Castelli, I. E., Mounet, N. & Marzari, N. Precision and efficiency in solid-state pseudopotential calculations. *npj Comput. Mater.* **4**, 72, (2018)

63. Eriksson, F., Fransson, E. & Erhart, P. The Hiphive Package for the Extraction of High-Order Force Constants by Machine Learning. *Adv. Theory Simulations* **2**, 1800184, (2019)

64. Lee, H. *et al.* Electron–phonon physics from first principles using the EPW code. *npj Comput. Mater.* **9**, 156, (2023)

65. Cepellotti, A., Coulter, J., Johansson, A., Fedorova, N. S. & Kozinsky, B. Phoebe: a high-performance framework for solving phonon and electron Boltzmann transport equations. *JPhys Mater.* **5**, 035003, (2022)

66. Carrete, J. *et al.* almaBTE : A solver of the space–time dependent Boltzmann transport equation for phonons in structured materials. *Comput. Phys. Commun.* **220**, 351, (2017)

67. Togo, A. & Seko, A. On-the-fly training of polynomial machine learning potentials in computing lattice thermal conductivity. *J. Chem. Phys.* **160**, 211001, (2024).

68. Tamura, S. I. Isotope scattering of dispersive phonons in Ge. *Phys. Rev. B* **27**, 858, (1983)

69. Pizzi, G. *et al.* Wannier90 as a community code: New features and applications. *J. Phys. Condens. Matter* **32**, 165902, (2020)

70. Marzari, N., Mostofi, A. A., Yates, J. R., Souza, I. & Vanderbilt, D. Maximally localized Wannier functions: Theory and applications. *Rev. Mod. Phys.* **84**, 1419, (2012)

**Acknowledgements**

Hio-Ieng Un, Henning Sirringhaus, Naoya Fukui, and Hiroshi Nishihara acknowledge the support from the Engineering and Physical Sciences Research Council (EPSRC) and the Japanese Society for the Promotion of Science (JSPS) through a core-to-core grant (EP/S030662/1). Hio-Ieng Un thank William Wood for discussion on Hall effect. Henning Sirringhaus thanks the Royal Society for a Research Professorship (RP/R1/201082). Ian Jacobs acknowledges funding from the European Research Council (Advanced Grant 101020872) and a Royal Society University Research Fellowship (URF\R1\231287). Jordi Ferrer Orri acknowledges funding from the Engineering and Physical Sciences Research Council (EPSRC) Nano Doctoral Training Centre (EP/L015978/1). David Cornil and David Beljonne acknowledge the support from the Energy Transition Fund of the Belgian Federal Government (FPS Economy) within the T-REX project. The computational resources for the Raman simulations were provided by the Consortium des "Equipements de Calcul Intensif" (CÉCI) funded by the Belgian National Fund for Scientific Research (F.R.S.-FNRS) under Grant 2.5020.11. David Beljonne is FNRS Research Director. SED studies were supported by the access to e02 at ePSIC Diamond Light Source (MG32017). We thank the Diamond Light Source Beamline I-07 for beamtime (SI35227, SI35227-1), and Jonathan Rawle for measurement assistance. We also thank the EPSRC and the Henry Royce Institute for access to the thermoelectric test equipment (Cambridge Royce facilities grant EP/P024947/1 and Sir Henry Royce Institute – recurrent grant EP/R00661X/1). Kamil Iwanowski acknowledges support from the Winton & Cavendish Scholarship at the Department of Physics in the University of Cambridge. Michele Simoncelli acknowledges support from Gonville and Caius College at the University of Cambridge. The computational resources for the first-principles simulations based on the Wigner transport equation were provided by: (i) the Kelvin2 HPC platform at the NI-HPC Centre (funded by EPSRC and jointly managed by Queen's University Belfast and Ulster University); (ii) the UK National Supercomputing Service ARCHER2, for which access was obtained via the UKCP consortium and funded by EPSRC (EP/X035891/1).

**Author contributions**

H.-I.U. conceived the project, prepared all samples and devices, designed and conducted all experimental characterizations with assistance from some of other authors except for the SED characterization which was done and analysed by J.F.O independently. H.-I.U analysed data, and led the scientific development of this work with H.S. and M.S. together. I.E.J. greatly helped with GIWAXS measurement and initial data processing of the raw data. J.F.O and I.E.J. reviewed and commented on the initial draft. N.F. synthesized the BHT molecule under H.N. supervision. D.C. and D.B. helped with the understanding on the Raman spectrum. K.I. and M.S. performed the theoretical analysis of the thermal conduction properties based on the Wigner Transport Equation, wrote the corresponding section in the paper, and contributed to the explanation of the experimental results. H.S. supervised the project. H.-I.U. wrote the first draft, and revised it with significant contribution from H.S. and inputs from M.S. All authors reviewed and commented on the manuscript.

## Competing interests

The authors declare no competing interests.